\pdfoutput=1

\documentclass[11pt,twoside,a4paper,cmspaper,final,collab]{cms-tdr}
\def\svnVersion{183695}\def\cmsCernNo{2013-033}\def\cmsCernDate{\today}\def\cmsMessage{Submitted to Physical Review Letters}

\begin{document}\cmsNoteHeader{TOP-12-014}

\hyphenation{had-ron-i-za-tion}
\hyphenation{cal-or-i-me-ter}
\hyphenation{de-vices}

\RCS$Revision: 179509 $
\RCS$HeadURL: svn+ssh://svn.cern.ch/reps/tdr2/papers/TOP-12-014/trunk/TOP-12-014.tex $
\RCS$Id: TOP-12-014.tex 179509 2013-04-04 08:12:35Z pandolf $
\newlength\cmsFigWidth
\ifthenelse{\boolean{cms@external}}{\setlength\cmsFigWidth{0.4\columnwidth}}{\setlength\cmsFigWidth{0.3\textwidth}}
\ifthenelse{\boolean{cms@external}}{\providecommand{\cmsLeft}{top}}{\providecommand{\cmsLeft}{left}}
\ifthenelse{\boolean{cms@external}}{\providecommand{\cmsRight}{bottom}}{\providecommand{\cmsRight}{right}}
\ifthenelse{\boolean{cms@external}}{\providecommand{\suppMaterial}{the supplemental material}}{\providecommand{\suppMaterial}{appendix~\ref{app:suppMat}}}
\cmsNoteHeader{TOP-12-014} 
\title{Measurement of associated production of vector bosons and \texorpdfstring{\ttbar in $\Pp\Pp$ collisions at $\sqrt{s}=7\TeV$}{t t-bar in pp collisions at sqrt{s}=7 TeV}}

\date{\today}

\abstract{
The first measurement of vector-boson production associated with a top quark-antiquark pair in proton-proton
collisions at $\sqrt{s} = 7\TeV$ is presented. The results are based on a dataset corresponding to an integrated luminosity of 5.0\fbinv, recorded
by the CMS detector at the LHC in 2011. The measurement is performed in two independent channels
through a trilepton analysis of $\ttbar \cPZ$ events and a same-sign dilepton analysis of $\ttbar V$ ($V=\PW\text{ or }\cPZ$) events.
In the trilepton channel a direct measurement of the $\ttbar \cPZ$ cross section
$\sigma_{\ttbar \cPZ} = 0.28^{+0.14}_{-0.11}\stat\; ^{+0.06}_{-0.03}\syst\unit{pb}$
is obtained.
In the dilepton channel a measurement of the $\ttbar V$ cross section yields
$\sigma_{\ttbar V} = 0.43^{+0.17}_{-0.15}\stat\; ^{+0.09}_{-0.07}\syst\unit{pb}$.
These measurements have a significance, respectively, of 3.3 and 3.0 standard deviations from the background hypotheses and are compatible,
within uncertainties, with the corresponding NLO predictions of $0.137^{+0.012}_{-0.016}$ and $0.306^{+0.031}_{-0.053}\unit{pb}$.
}

\hypersetup{%
pdfauthor={CMS Collaboration},%
pdftitle={Measurement of associated production of vector bosons and t t-bar at in pp collisions sqrt(s)= 7 TeV},%
pdfsubject={CMS},%
pdfkeywords={CMS, physics, top}}

\maketitle

\begin{figure}[b]
\centering
\includegraphics[width=\cmsFigWidth]{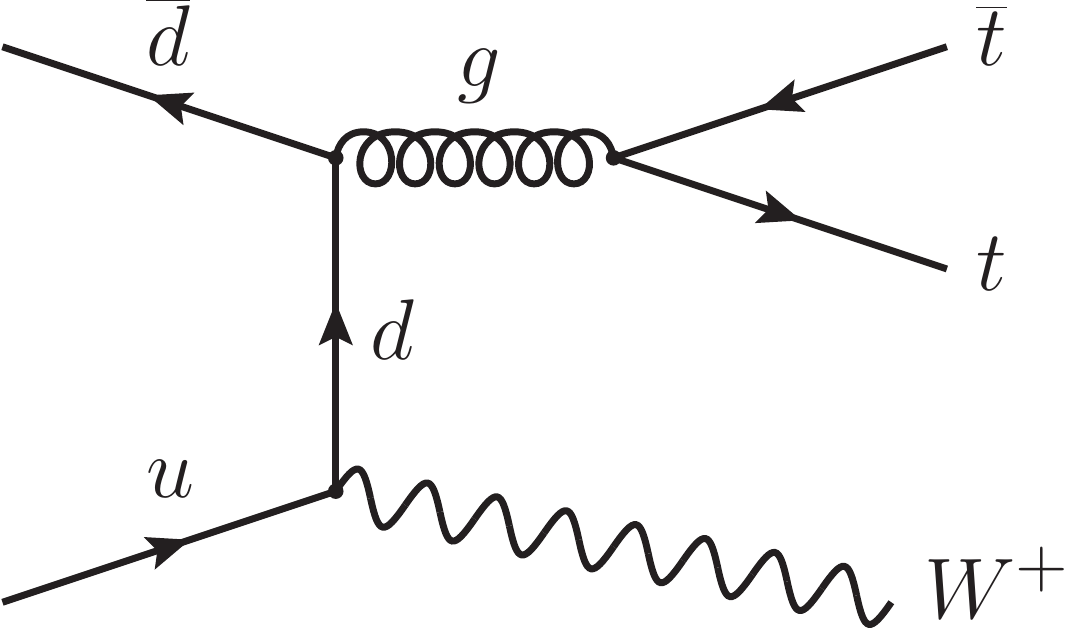}\hfil
\includegraphics[width=\cmsFigWidth]{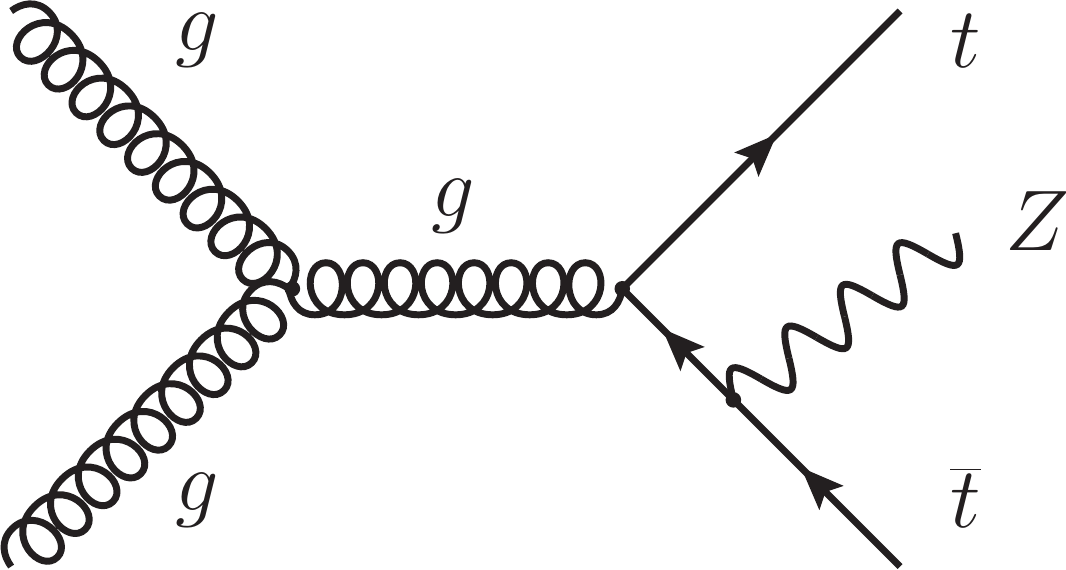}
\caption{Most important leading-order Feynman diagrams for $\ttbar\PW$ and $\ttbar \cPZ$ production in proton-proton collisions. The charge conjugate of the diagrams shown is implied.\label{fig:diagrams}}
\end{figure}

Although the top quark was discovered more than 15 years ago~\cite{top_cdf, top_d0_discovery}, many of its properties have not yet been fully investigated.
In particular, most of its couplings have never been directly measured.
The large value of the its mass indicates that the top quark could play a special role in the context of electroweak symmetry breaking.
Extensions of the standard model (SM), such as technicolor or other scenarios with a strongly coupled Higgs sector, could alter the top-quark couplings.
A measurement of the production of a top-quark pair in association with vector bosons is a key test of the validity of the SM at the\TeV scale.
In Fig.~\ref{fig:diagrams} the most important leading-order Feynman diagrams for $\ttbar\PW$ and $\ttbar \cPZ$ production in proton-proton collisions
are shown.
The current estimate of the cross section for these processes is based on quantum chromodynamics~(QCD) calculations at next-to-leading order (NLO),
which yield $0.169^{+0.029}_{-0.051}\unit{pb}$~\cite{ttWxsecNLO}
for $\ttbar\PW$ production, and $0.137^{+0.012}_{-0.016}\unit{pb}$~\cite{ttZpubNEWER} for $\ttbar \cPZ$ production.

In this Letter, the first measurement of the cross section for associated production of a vector boson and a $\ttbar$ pair is presented.
Two  analyses are conducted: one based on trilepton signatures
produced in $\ttbar \cPZ$ decays, and one based on same-sign dilepton signatures produced by
 $\ttbar V$ events (with $V=\PW$ or $\cPZ$).

This measurement uses data from proton-proton collisions, produced at a
center-of-mass energy of 7\TeV, corresponding to an integrated luminosity of $5.0 \pm 0.1\fbinv$~\cite{CMS:lumi}. The data
were collected by the Compact Muon Solenoid (CMS) detector at the Large Hadron Collider (LHC) in 2011.
As the signal would appear as an excess over a background of similar size,
the background estimation is a focus of the analysis.
The majority of background contributions are estimated using the data,
while remaining background processes are estimated using Monte Carlo (MC) simulations.
Simulated MC event samples are generated using the \MADGRAPH~5.1.3.30 event generator~\cite{MADGRAPH}, interfaced with \PYTHIA~6.4~\cite{PYTHIA} for parton showering.
The same generator chain is used for signal events. A \textsc{Geant4}-based~\cite{GEANT4} simulation of the response of the CMS detector is used for both signal and background events.
These events are processed with the same reconstruction algorithms as the data. Simulated event yields are scaled to
the integrated luminosity in the data using cross section calculations to the highest order available, taking into account the trigger and reconstruction efficiencies determined from the data.
In addition, the simulated distribution of the number of simultaneous proton-proton collisions within the same bunch crossing (pileup) is reweighted to match the one observed in the data.

A detailed description of the CMS detector can be found elsewhere~\cite{CMS:2010}.
Its central feature is a 3.8\unit{T} superconducting solenoid of 6\unit{m} internal diameter.
Within its field volume are the silicon tracker, the crystal electromagnetic calorimeter (ECAL), and the brass/scintillator sampling hadron calorimeter (HCAL).
The muon system, composed of drift tubes, cathode strip chambers, and resistive-plate chambers, is installed outside the solenoid, embedded in the steel return yoke.
CMS uses a right-handed coordinate system, with the origin at the nominal interaction point, the $x$ axis pointing to the center of the LHC, the $y$ axis pointing up (perpendicular to the LHC plane), and the $z$ axis along the counterclockwise-beam direction.
The polar angle $\theta$ is measured from the positive $z$ axis and the azimuthal angle $\phi$ is measured in the $x$-$y$ plane.
The pseudorapidity $\eta$ is defined as $-\ln[ \tan(\theta/2) ]$.

Muons~\cite{muons} are measured with the combination of the tracker and the muon system, in the pseudorapidity range $|\eta|<2.4$.
Electrons~\cite{electrons} are detected as tracks in the tracker pointing to energy clusters in the ECAL up to $|\eta|= 2.5$.
Both muons and electrons are required to have a momentum transverse to the beam axis, \pt, greater than 20\GeV.
Both the $\pt$ and $\eta$ requirements are consistent with those employed in the online trigger selection, where the presence of two
isolated charged leptons, either electrons or muons, in any flavor combination, is required to accept the events.

 The full details of electron and muon identification criteria are described elsewhere~\cite{VBTF}.
Isolation requirements on lepton candidates are enforced by measuring the additional detector activity in a surrounding
cone of $\Delta R \equiv \sqrt{(\Delta\eta)^2+(\Delta\phi)^2} <0.3$, where $\Delta\eta$ and $\Delta\phi$ are
the differences in pseudorapidity and in azimuthal angle, measured in radians, respectively.
For muons the total sum of the transverse momenta of the additional reconstructed tracks and of the energy in the
calorimeters in the surrounding cone is required to be less than 15\% of the muon transverse momentum in the trilepton channel and 5\%
in the dilepton channel.
 Electron isolation requirements are similar but vary depending on the shape of the electron shower.
To minimize the contribution of lepton candidates arising from jet misidentification, tighter isolation and
identification requirements are employed in the dilepton channels.

Jets are reconstructed with a particle-flow (PF) algorithm~\cite{CMS-PAS-PFT-09-001}, a global event reconstruction technique
which optimally combines the information of all sub-detectors to reconstruct the particles produced in a collision.
Reconstructed particle candidates are clustered to form PF jets with the anti-\kt algorithm~\cite{Cacciari:2008gp} with a distance parameter of 0.5.
The jet energy resolution is typically 15\% at 10\GeV and 8\% at 100\GeV.
Jets are required to be inside the tracker acceptance ($|\eta|<2.4$), to increase the reconstruction
efficiency and the precision of the energy measurement using PF techniques.
Jet energy corrections are applied to account for the non-linear response of the calorimeters to the particle energies and other instrumental effects.
These corrections are based on in situ measurements using dijet and $\gamma+\text{jet}$ data samples~\cite{jes}.
Pileup activity has an effect on jet reconstruction by contributing additional particles to the reconstructed jets.
The average energy density due to pileup is evaluated in each event and the corresponding energy is subtracted from each jet~\cite{Cacciari:2008gn}.
A jet identification requirement, primarily based on the energy balance between charged and neutral hadrons in a jet, is applied to remove misidentified jets.
 Jets are required to have $\pt > 20$\GeV.

To identify jets originating from the hadronization of bottom quarks, a \cPqb-tagging algorithm~\cite{CMS-BTV-12-001} is employed.
The algorithm identifies jets from \cPqb-hadron decays by
requiring at least two tracks to have significant impact parameters with respect to the primary interaction vertex. This tagger is used here with two operating points:
the \emph{loose} point corresponds to an efficiency for jets originating from $\cPqb$ quarks of about 80\%
and a misidentification probability for jets from light quarks and gluons of 10\%, while
the \emph{medium} operating point provides an efficiency for $\cPqb$ jets of about 65\%  and a misidentification probability of about 1\%.

In the trilepton analysis, events originating from the process
\ifthenelse{\boolean{cms@external}}{
\begin{linenomath}
\begin{multline*}
\Pp\Pp\to \ttbar \cPZ \rightarrow  (\cPqt\to \cPqb\ell^{\pm}\nu)(\cPqt\to \cPqb jj)(\cPZ \to \ell^{\pm}\ell^{\mp}) \\ (\text{with } \ell = \Pe \text{ or } \mu )
\end{multline*}
\end{linenomath}
}
{
\begin{linenomath}
\begin{equation*}
\Pp\Pp\to \ttbar \cPZ \rightarrow  (\cPqt\to \cPqb\ell^{\pm}\nu)(\cPqt\to \cPqb jj)(\cPZ \to \ell^{\pm}\ell^{\mp}) \quad\quad (\text{with } \ell = \Pe \text{ or } \mu )
\end{equation*}
\end{linenomath}
}
are selected if they contain two same-flavor, opposite-charge leptons (electrons or muons) with $\pt > 20$\GeV,
where the dilepton system must have an invariant mass between 81 and 101\GeV and $\pt > 35$\GeV.
The presence of a third lepton with $\pt > 10$\GeV and at least three jets,
two of which are positively \cPqb-tagged~(one medium and one loose tag), is required, and the scalar sum of the $\pt$
of all selected jets~(\HT) is required to be larger than 120\GeV.
These selection requirements have been chosen by optimizing the expected significance of the measurement.

The main background contributions in this analysis are dilepton events from the Drell--Yan process and from $\ttbar$ events,
where a  third lepton is reconstructed from hadronization products, and $\PW\cPZ$ events where both vector bosons decay to leptons.
To determine the background contributions from the data, event samples with less stringent requirements are used.
Dilepton $\Pe\mu$ events which satisfy only the lepton $\pt$ and jet multiplicity requirements are
dominated by top-quark pair production, and are used to control the normalization of the $\ttbar$ simulation. A
normalization factor of $1.05 \pm 0.12$ with respect to the NLO cross section is found.
The normalization of the Drell--Yan and $\PW\cPZ$ simulations is determined from a control sample where
all the signal requirements are met, except there are no \cPqb-tagged jets.
The simulations must be normalized by a factor of $1.30 \pm 0.13$  to correctly predict the number of events in
the \cPZ-mass peak in this background-dominated region.
Sources of background arising from single-top-quark production mediated through a virtual $\PW$ boson, in conjunction with
a $\cPZ$ boson ($\cPqt\cPqb\cPZ$), are taken from the simulation, scaled to the leading order
cross section, and an uncertainty of 50\% is assumed on this yield.
The contribution from events containing a SM Higgs boson, assuming a mass of 125\GeV, as suggested by recent findings~\cite{cms_higgs, atlas_higgs}, has been estimated and found negligible for the trilepton channel.

The total systematic uncertainty is evaluated by assessing the relative change in signal efficiency and
background yield in the simulation when varying relevant parameters by one standard deviation.
The sources of systematic uncertainty include experimental uncertainties such as the background estimate, lepton reconstruction
and trigger efficiencies, jet energy scale and resolution, \cPqb-tagging efficiency, pileup modeling, and the integrated luminosity.
Model uncertainties arising from scale variations of the
matrix-element/parton-shower matching scale and the hard-scattering scale $Q^2$ are also included. The dominant uncertainty comes from
the background estimate and amounts to 27\% of the background yield; this includes the statistical uncertainty on the number of simulated events and the uncertainty on the
background scale factors determined from the data all added in quadrature. All other uncertainties are less than 5\%.
The signal efficiencies are determined from MC simulations using \MADGRAPH.
In order to account for any difference due to the NLO predictions, signal efficiencies are also calculated using the \POWHEG~\textsc{box}~\cite{Nason:2004rx,Frixione:2007vw,Alioli:2010xd} generator.
The two simulations differ in their predictions of the signal efficiencies by 13\%,
and this value is taken as a systematic uncertainty.
Systematic uncertainties that affect both signal and background yields are assumed to be fully correlated.
The total systematic uncertainty on the measured cross section is 15\%.

\begin{figure}[t]
\centering
\includegraphics[width=0.38\textwidth]{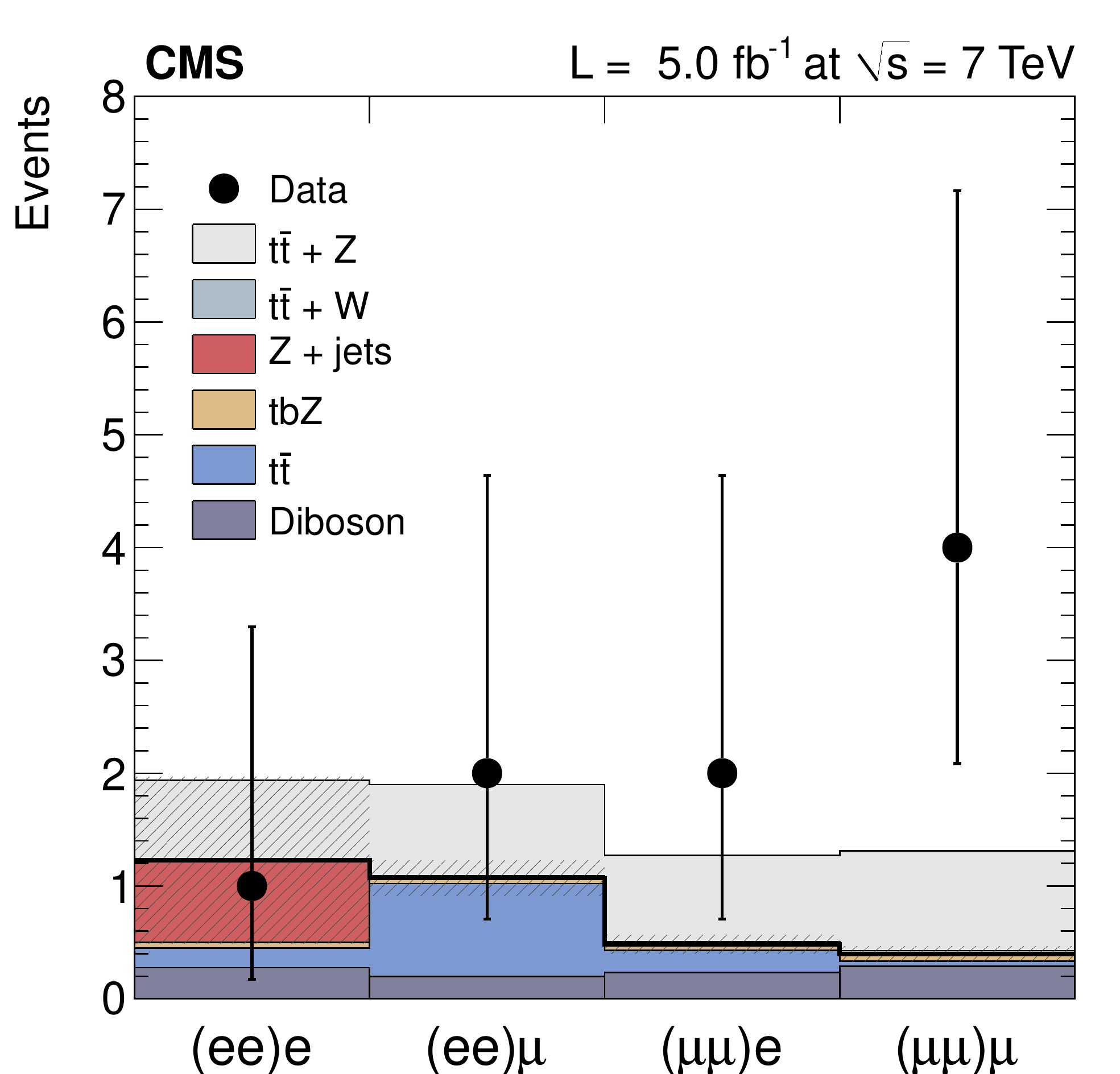}
\includegraphics[width=0.38\textwidth]{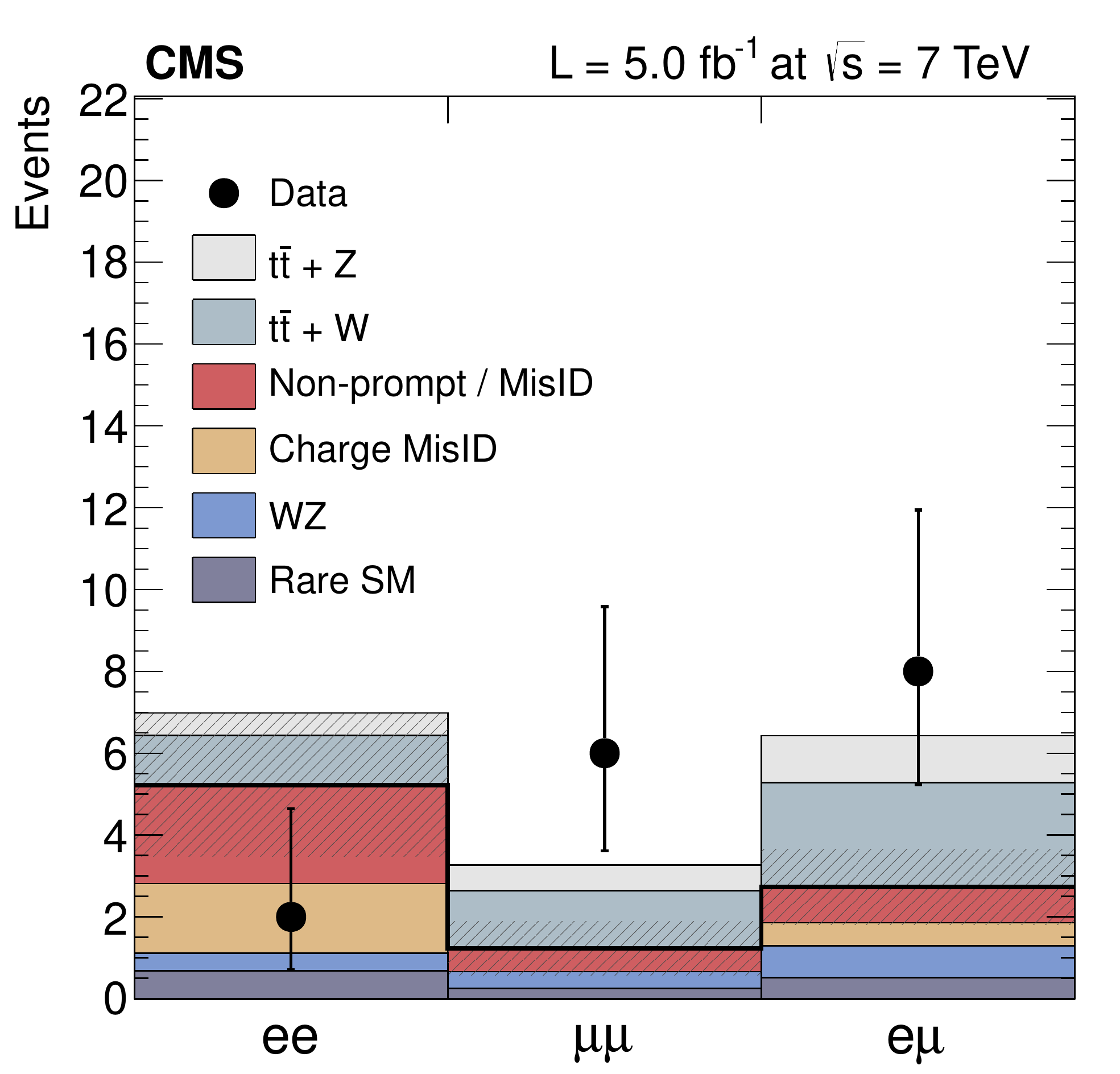}
\caption{Event yields after final selection requirements, separated in lepton flavor channels for the
trilepton~(\cmsLeft) and same-sign dilepton~(\cmsRight) analyses.
The expected contributions from signal and background processes are shown, and the uncertainty on the estimated background yield is superimposed with a
grey hashed band. \label{fig:yields}}
\end{figure}

The event yields after applying the full event selection are shown in Fig.~\ref{fig:yields}.
Nine events are observed, compared to a background expectation of  $3.2 \pm 0.8$ events.
From the combination of the four decay channels,
 the presence of a $\ttbar \cPZ$ signal is established
with a combined significance of 3.3 standard deviations, corresponding to a $p$-value of $4\times 10^{-4}$,
as obtained with an asymptotic profile likelihood estimator~\cite{profileLike}.
The cross section is extracted through a simultaneous measurement performed in the four decay channels,
and is measured to be
\begin{linenomath}
\begin{equation*}
\sigma_{\ttbar \cPZ} = 0.28^{+0.14}_{-0.11}\stat\;^{+0.06}_{-0.03}\syst\unit{pb.}
\end{equation*}
\end{linenomath}
The measured cross section is found to be compatible, within uncertainties, with the NLO prediction of $0.137^{+0.012}_{-0.016}\unit{pb}$~\cite{ttZpubNEWER}.
A comparison of the observed and predicted distributions for several kinematic variables is available in \suppMaterial.

The same-sign dilepton analysis searches for events with the following decay chains:
\ifthenelse{\boolean{cms@external}}{
\begin{align*}
\Pp\Pp&\to\ttbar\PW \to  (\cPqt\to \cPqb\ell^{\pm}\nu)(\cPqt\to \cPqb jj)(\PW \to \ell^{\pm}\nu); \\
\Pp\Pp&\to\ttbar\cPZ \to (\cPqt\to \cPqb\ell^{\pm}\nu)(\cPqt\to \cPqb jj)(\cPZ \to \ell^{\pm}\ell^{\mp})\\
&\omit\hfill$\displaystyle{(\text{with } \ell = \Pe\text{ or }\mu)}$.
\end{align*}
}
{
\begin{linenomath}
\begin{align*}
\Pp\Pp&\to\ttbar\PW \to  (\cPqt\to \cPqb\ell^{\pm}\nu)(\cPqt\to \cPqb jj)(\PW \to \ell^{\pm}\nu); \\
\Pp\Pp&\to\ttbar\cPZ \to (\cPqt\to \cPqb\ell^{\pm}\nu)(\cPqt\to \cPqb jj)(\cPZ \to \ell^{\pm}\ell^{\mp}) \quad\quad (\text{with } \ell = \Pe\text{ or }\mu).
\end{align*}
\end{linenomath}
}

The final set of selection criteria for the dilepton channel requires the presence of
two same-sign leptons, one with $\pt > 55$ and the other with $\pt > 30$\GeV,
and a dilepton invariant mass greater than 8\GeV,
at least three jets with $\pt > 20$\GeV of which at least one is \cPqb-tagged by the medium operating point,
and $\HT > 100$\GeV. These selection requirements have been chosen by optimizing the expected significance of the signal excess.
To make this data sample statistically independent of the data selected for the trilepton channel,
events passing the trilepton selection are removed.

The benefit of searching for same-sign dilepton events is that SM processes containing two prompt same-sign
leptons in the final state have very small cross sections.
The background processes considered here include diboson production ($\PW\cPZ$, $\cPZ\cPZ$, $\PW\gamma$, $\cPZ\gamma$, $\PW^\pm \PW^\pm$),
$\cPqt\cPqb\cPZ$, triboson production, and  production of vector-boson pairs from double-parton scattering.
Yields from these processes are taken directly from the simulation and scaled to NLO predictions whenever available.

The dominant background contributions originate from non-prompt leptons or misreconstruction effects:
pions in jets or decay products of heavy-flavor mesons may give rise to non-prompt lepton candidates;
charge misidentification in events with opposite-sign lepton pairs results in same-sign events.
These background rates are determined from control regions in the data using
techniques that determine the prompt and non-prompt lepton misidentification rates from QCD dijet and $\cPZ \rightarrow \ell\ell$ event samples~\cite{SUS-11-010}.
The result is an estimate, fully based on control samples in the data, of backgrounds with one or more lepton candidates that are not reconstructed from a prompt final-state lepton.
These include semi-leptonic
$\ttbar$ decays, Drell--Yan events with hard jet production, and QCD multijet production.

The background estimate due to charge misidentification of one of the leptons is obtained from the number of opposite-sign dilepton events
in the signal region and the probability to wrongly measure the charge of a lepton.
This probability is negligible for muons, but considerable for electrons.
From the fraction of same-sign events in a control region dominated by $\cPZ$ decay, the electron charge misidentification
probability is measured to be 0.02\% (0.3\%) in the barrel (endcap) region of the detector.

Systematic uncertainties relative to experimental measurements or model uncertainties
are evaluated in a similar manner as in the trilepton channel, and are expressed
in terms of uncertainties on the signal efficiency or the background yield.
Uncertainties on the background prediction are quantified differently for
each of the background yield estimates: a 50\% uncertainty is assigned to the
estimate of processes with non-prompt leptons;
the uncertainty on charge misidentification backgrounds is driven by the uncertainty on the measured single-lepton charge misidentification probability
and amounts to about 20\%; the uncertainty on $\PW\cPZ$ production is taken from the CMS cross section measurement and
is equal to 20\%; for all the other SM processes taken from simulation, most of which have not been measured yet, an uncertainty of 50\% is assigned.
Similar to the trilepton analysis, the uncertainty of the signal efficiency is estimated to be 13\%.
All uncertainties that affect both signal and background yields are assumed to be fully
correlated, whereas background prediction uncertainties are uncorrelated.
The total systematic uncertainty in the dilepton channel is 15\%.
The contribution from a SM Higgs boson with a mass of 125\GeV to the same-sign dilepton sample is estimated to be as large as 0.8 events.
The majority of these events originate from Higgs boson production in associated production with $\ttbar$ pairs,
in conjunction with the decay channels $\PH\rightarrow \PW\PW$ and $\PH \rightarrow \tau\tau$.
This contribution is not included in the background estimation for this analysis, as doing so would assume a degree of knowledge about the SM Higgs which has not been verified yet.

Signal and background event yields are obtained as
shown in Fig.~\ref{fig:yields}.
A total of 16 events is selected in the data, compared to an expected background contribution of $9.2\pm 2.6$ events.
The presence of a $\ttbar V$ ($V=\PW$ or \cPZ) signal is established with a significance equivalent to 3.0 standard deviations
and a corresponding $p$-value of 0.002, as computed by multiplying the likelihoods of the three decay channels with an
asymptotic profile likelihood estimator.
 The combined cross section, as measured simultaneously from the three channels, is
\begin{linenomath}
\begin{equation*}
\sigma_{\ttbar V} = 0.43^{+0.17}_{-0.15}\stat\;^{+0.09}_{-0.07}\syst\unit{pb}.
\end{equation*}
\end{linenomath}
The measured cross section is compatible with the NLO prediction of $0.306^{+0.031}_{-0.053}\unit{pb}$.
A comparison of the observed and predicted distributions for several kinematic variables is available in \suppMaterial.

\begin{figure}[t]
	\centering
	\includegraphics[width=0.5\textwidth]{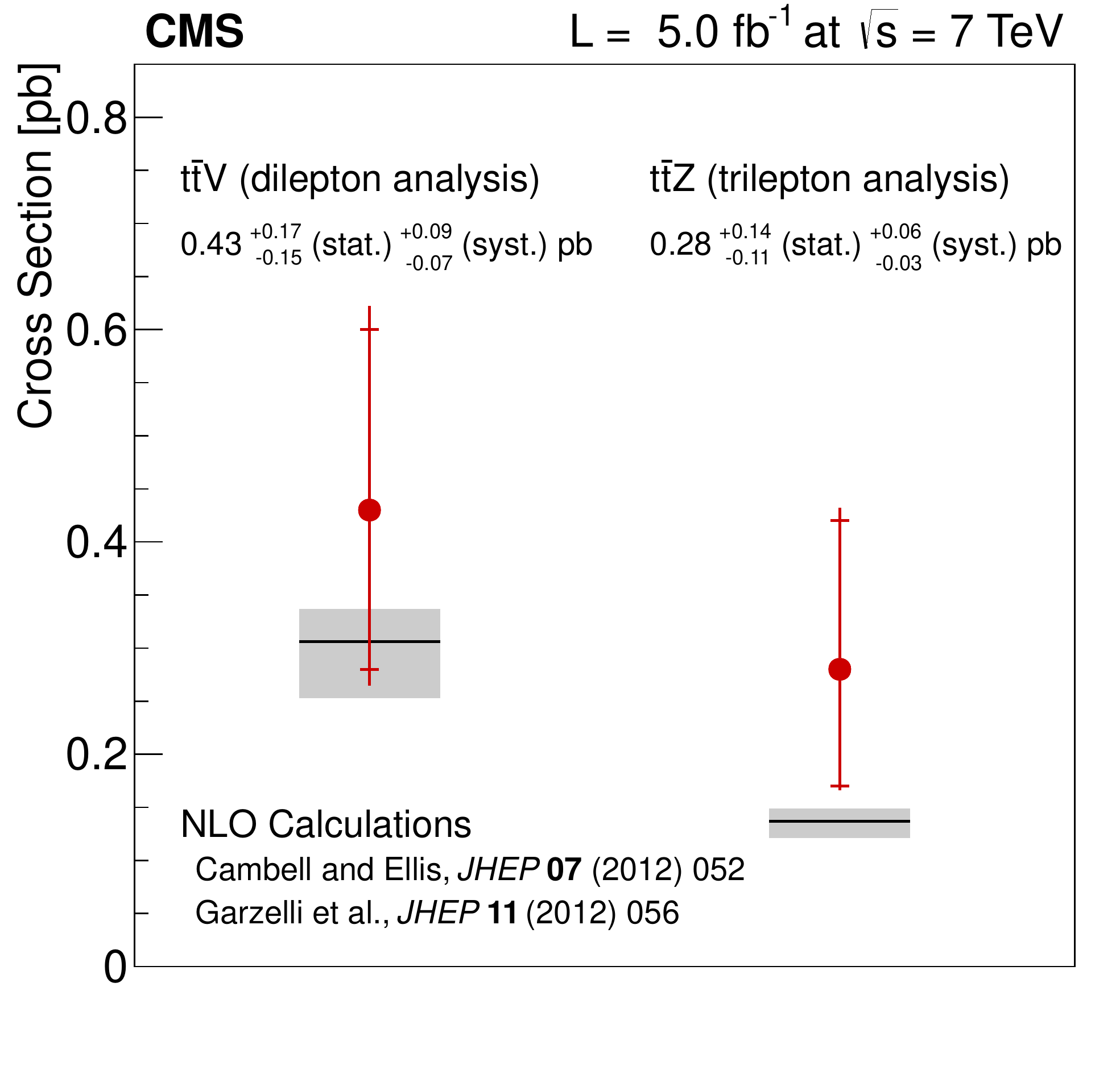}
	\caption{Measurements of the $\ttbar \cPZ$  and $\ttbar V$ production cross sections,
in the same-sign dilepton~(left) and trilepton channel~(right), respectively. The measurements are compared to the NLO calculations (horizontal
black lines) and their uncertainty (grey bands).
Internal error bars for the measurements represent the statistical component of the uncertainty.
}
	\label{fig:measurement_vs_NLO}
\end{figure}

In summary, the first measurement of the cross section of vector boson production associated with a top quark-antiquark pair at $\sqrt{s}=7\TeV$ has been presented.
In the trilepton channel a direct measurement of the $\ttbar \cPZ$ cross section
$\sigma_{\ttbar \cPZ} = 0.28^{+0.14}_{-0.11}\stat\;^{+0.06}_{-0.03}\syst\unit{pb}$
is obtained, with a significance of 3.3 standard deviations from the background hypothesis.
In the dilepton channel a measurement of the $\ttbar V$ process yields
$\sigma_{\ttbar V} = 0.43^{+0.17}_{-0.15}\stat\;^{+0.09}_{-0.07}\syst\unit{pb},$
with a significance of 3.0 standard deviations from the background hypothesis.
Both cross section measurements are compatible with the NLO predictions.
These results are summarized in Fig.~\ref{fig:measurement_vs_NLO}.

\section*{Acknowledgements}
We congratulate our colleagues in the CERN accelerator departments for the excellent performance of the LHC and thank the technical and administrative staffs at CERN and at other CMS institutes for their contributions to the success of the CMS effort. In addition, we gratefully acknowledge the computing centres and personnel of the Worldwide LHC Computing Grid for delivering so effectively the computing infrastructure essential to our analyses. Finally, we acknowledge the enduring support for the construction and operation of the LHC and the CMS detector provided by the following funding agencies: BMWF and FWF (Austria); FNRS and FWO (Belgium); CNPq, CAPES, FAPERJ, and FAPESP (Brazil); MEYS (Bulgaria); CERN; CAS, MoST, and NSFC (China); COLCIENCIAS (Colombia); MSES (Croatia); RPF (Cyprus); MoER, SF0690030s09 and ERDF (Estonia); Academy of Finland, MEC, and HIP (Finland); CEA and CNRS/IN2P3 (France); BMBF, DFG, and HGF (Germany); GSRT (Greece); OTKA and NKTH (Hungary); DAE and DST (India); IPM (Iran); SFI (Ireland); INFN (Italy); NRF and WCU (Republic of Korea); LAS (Lithuania); CINVESTAV, CONACYT, SEP, and UASLP-FAI (Mexico); MSI (New Zealand); PAEC (Pakistan); MSHE and NSC (Poland); FCT (Portugal); JINR (Armenia, Belarus, Georgia, Ukraine, Uzbekistan); MON, RosAtom, RAS and RFBR (Russia); MSTD (Serbia); SEIDI and CPAN (Spain); Swiss Funding Agencies (Switzerland); NSC (Taipei); ThEPCenter, IPST and NSTDA (Thailand); TUBITAK and TAEK (Turkey); NASU (Ukraine); STFC (United Kingdom); DOE and NSF (USA).

\bibliography{auto_generated}   
\ifthenelse{\boolean{cms@external}}{}{
\clearpage
\appendix
\section{Distributions of kinematic variables after final selection\label{app:suppMat}}
Comparisons are provided between the data and the signal and background predictions,
for events passing the full set of selection requirements. In Fig.~\ref{fig:trilepton_kinematics}
the comparisons are shown for the trilepton channel: the transverse momentum \pt(\cPZ) of the $\cPZ$ boson~(left), the scalar sum $\HT$ of all jet transverse momenta~(center),
and the missing transverse energy \ETslash, computed as the norm of the vectorial sum
of all particle-flow candidates in the event~(right).
Fig.~\ref{fig:ssdl_kinematics} shows the comparisons for the same-sign dilepton channel:
leading lepton transverse momentum~(left), $\HT$~(center) and \ETslash~(right).
\begin{figure*}[tbhp]
\centering
\includegraphics[width=0.32\textwidth]{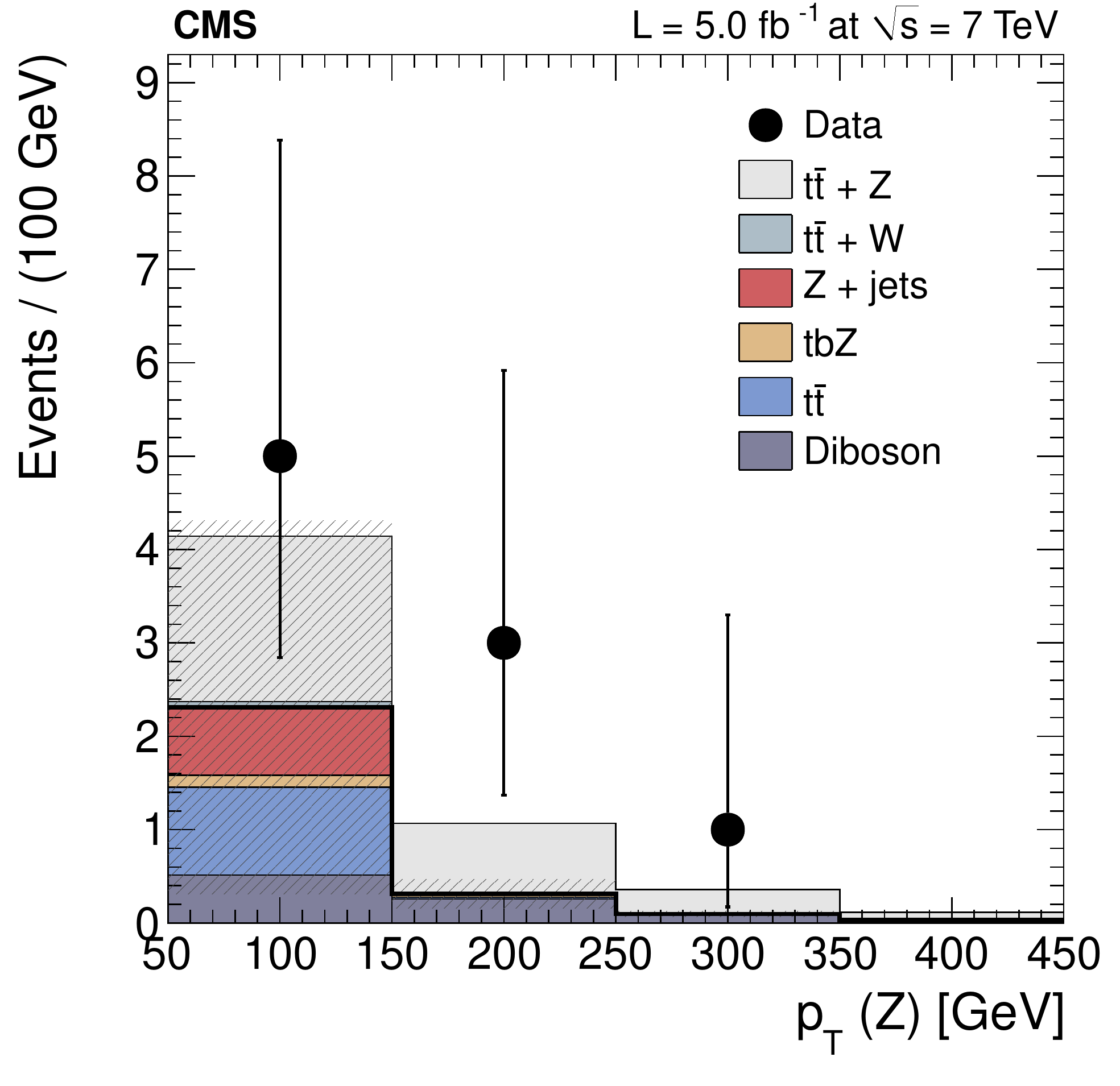}
\includegraphics[width=0.32\textwidth]{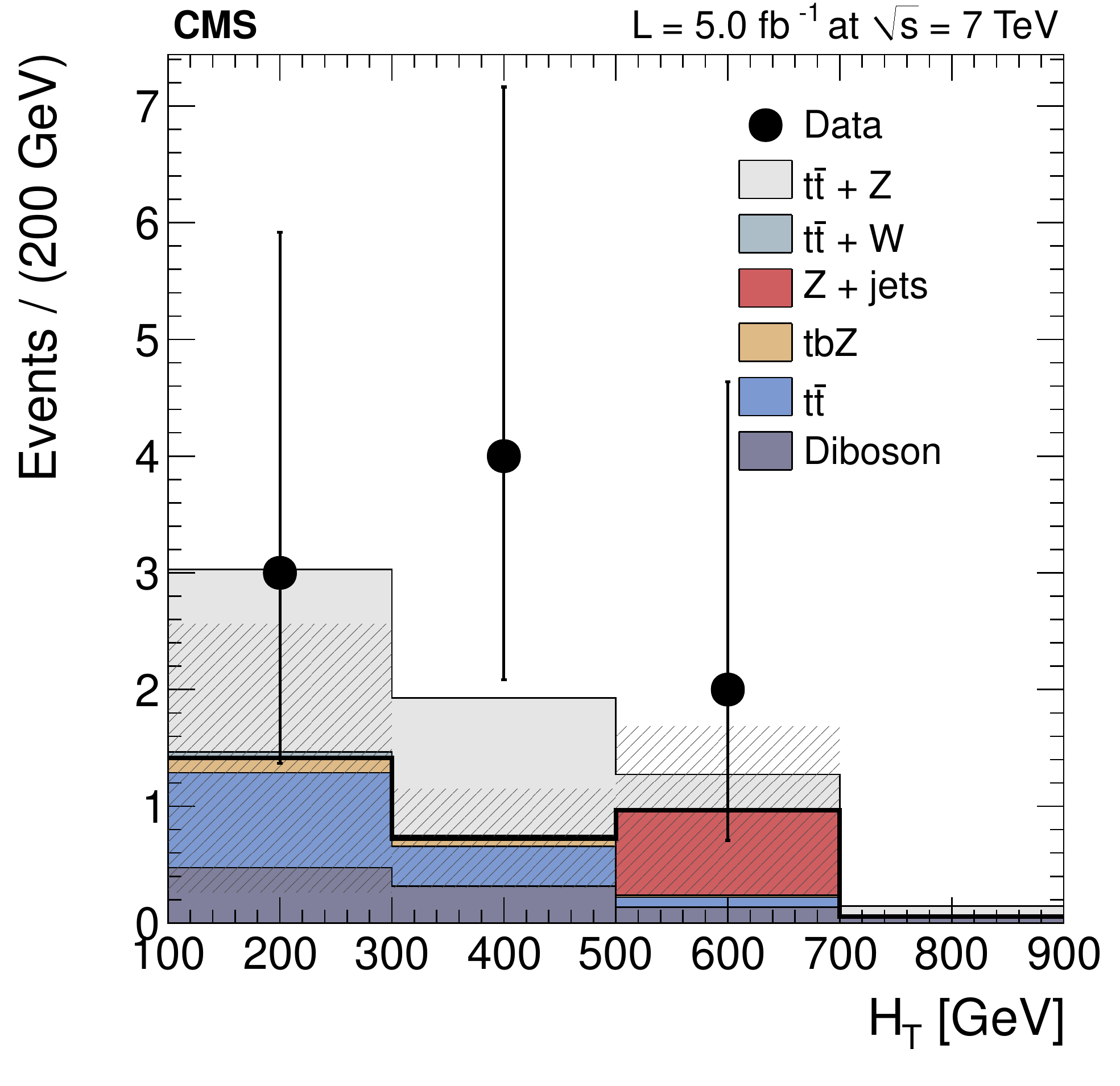}
\includegraphics[width=0.32\textwidth]{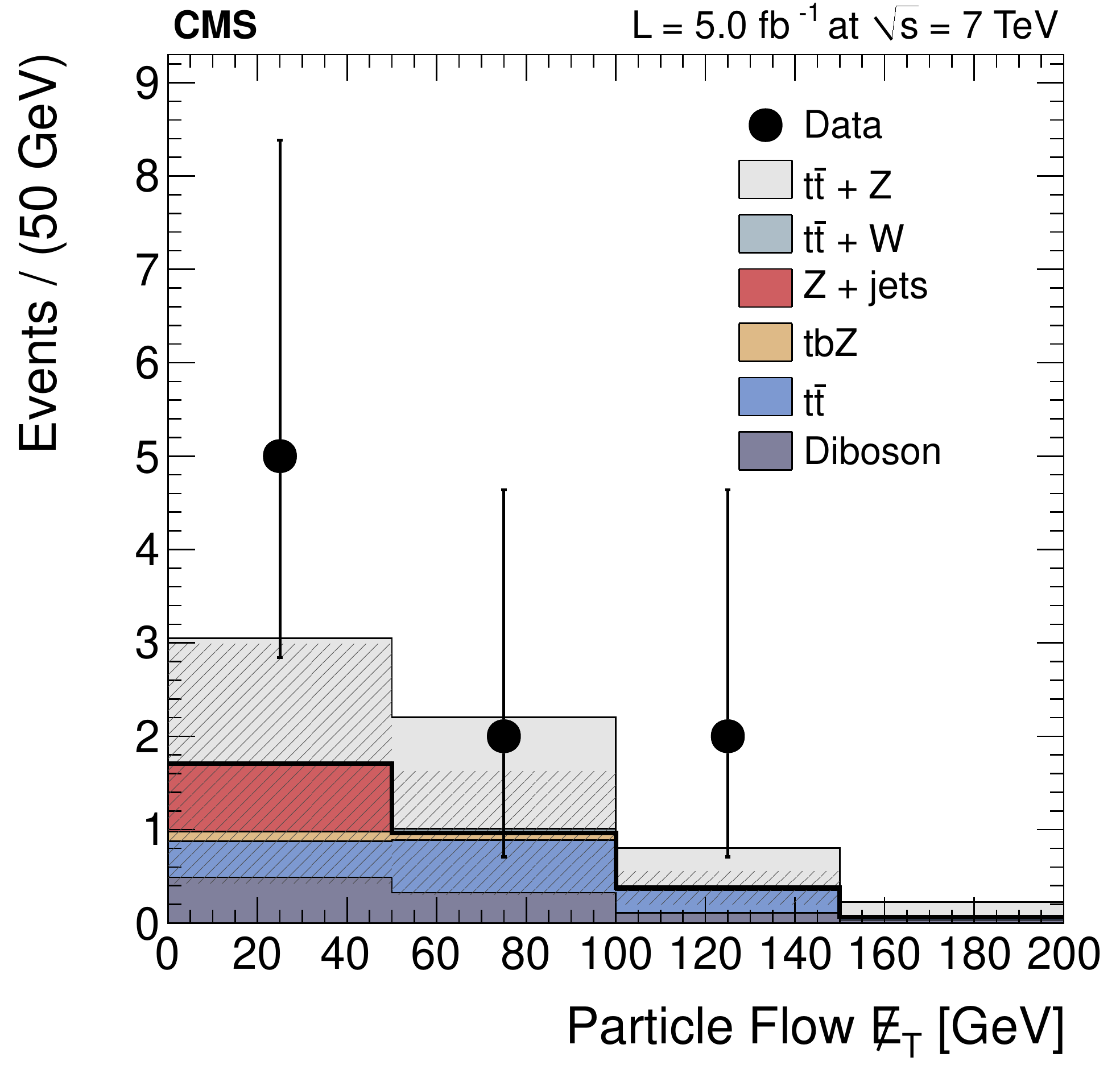}
\caption{Distributions of kinematic variables after final selection requirements for the trilepton analysis:
transverse momentum of the $\cPZ$ boson~(left), $\HT$~(center) and \ETslash~(right).
The uncertainty on the background yield is superimposed with a grey hashed band.
\label{fig:trilepton_kinematics}}
\end{figure*}
\begin{figure*}[tbhp]
\centering
\includegraphics[width=0.32\textwidth]{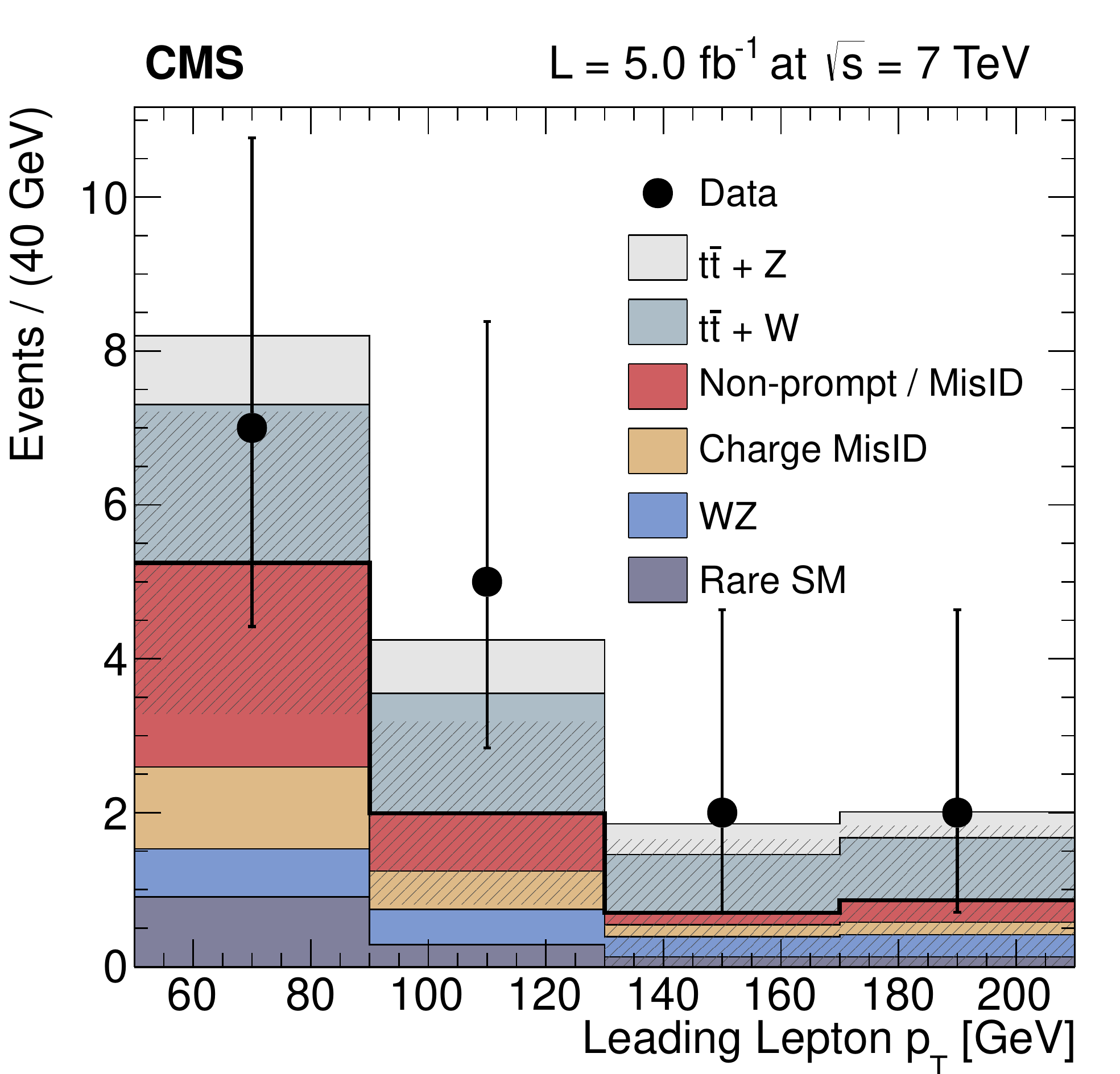}
\includegraphics[width=0.32\textwidth]{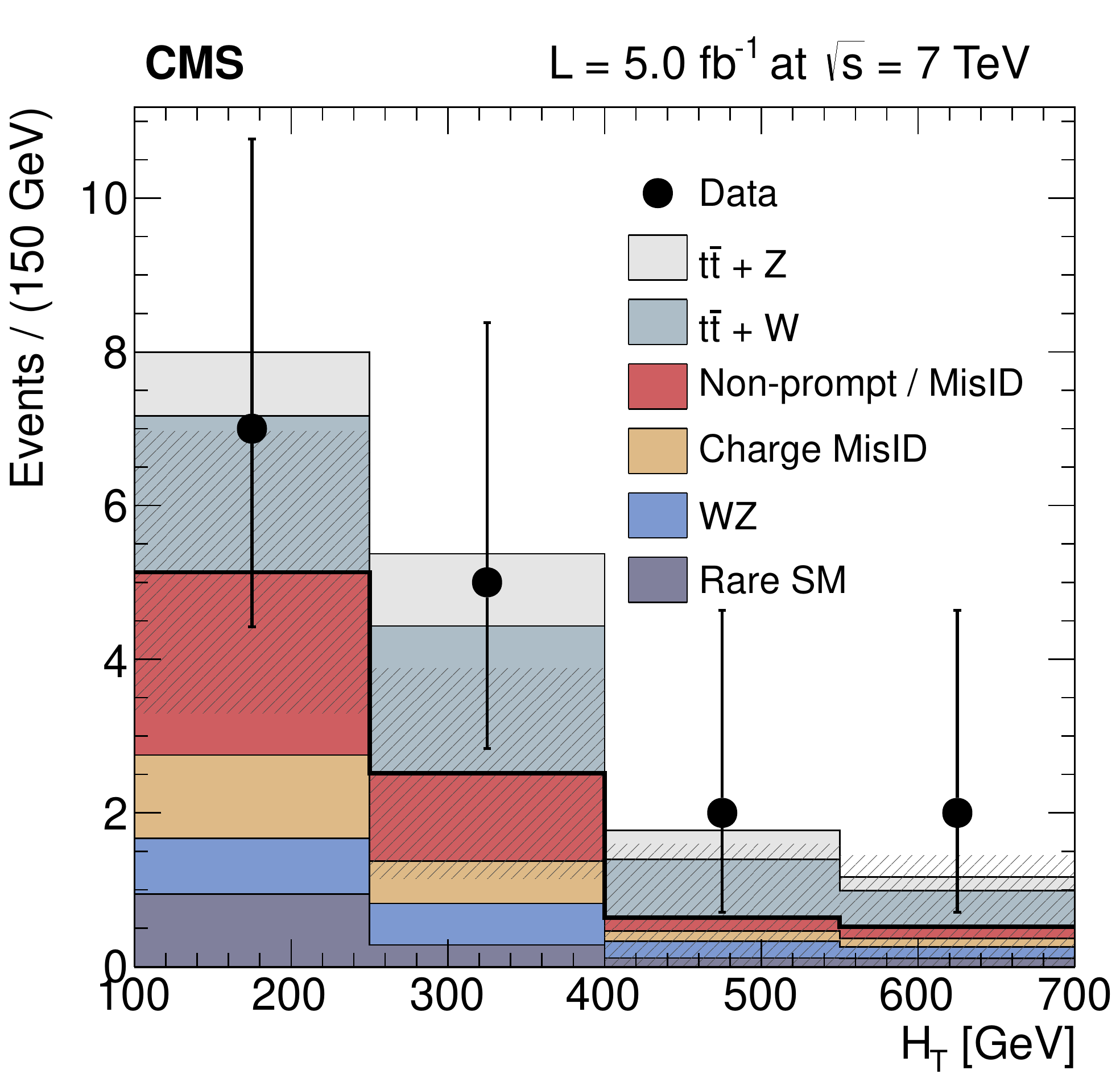}
\includegraphics[width=0.32\textwidth]{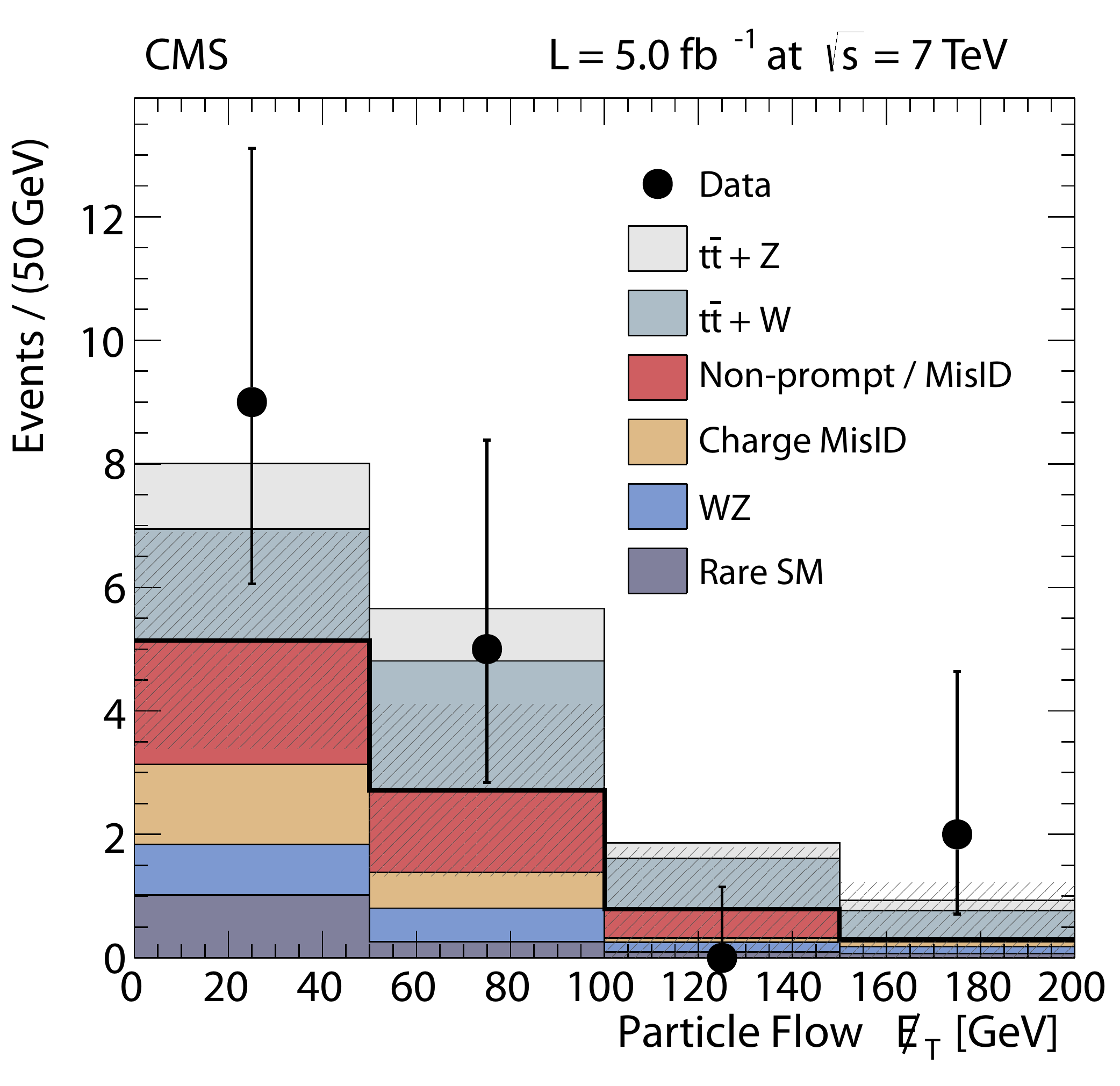}
\caption{Distributions of kinematic variables after final selection requirements for the same-sign dilepton analysis:
leading lepton transverse momentum~(left), $\HT$~(center) and \ETslash~(right).
The uncertainty on the background yield is superimposed with a grey hashed band.
\label{fig:ssdl_kinematics}}
\end{figure*}

}

\cleardoublepage \appendix\section{The CMS Collaboration \label{app:collab}}\begin{sloppypar}\hyphenpenalty=5000\widowpenalty=500\clubpenalty=5000\textbf{Yerevan Physics Institute,  Yerevan,  Armenia}\\*[0pt]
S.~Chatrchyan, V.~Khachatryan, A.M.~Sirunyan, A.~Tumasyan
\vskip\cmsinstskip
\textbf{Institut f\"{u}r Hochenergiephysik der OeAW,  Wien,  Austria}\\*[0pt]
W.~Adam, E.~Aguilo, T.~Bergauer, M.~Dragicevic, J.~Er\"{o}, C.~Fabjan\cmsAuthorMark{1}, M.~Friedl, R.~Fr\"{u}hwirth\cmsAuthorMark{1}, V.M.~Ghete, N.~H\"{o}rmann, J.~Hrubec, M.~Jeitler\cmsAuthorMark{1}, W.~Kiesenhofer, V.~Kn\"{u}nz, M.~Krammer\cmsAuthorMark{1}, I.~Kr\"{a}tschmer, D.~Liko, I.~Mikulec, M.~Pernicka$^{\textrm{\dag}}$, D.~Rabady\cmsAuthorMark{2}, B.~Rahbaran, C.~Rohringer, H.~Rohringer, R.~Sch\"{o}fbeck, J.~Strauss, A.~Taurok, W.~Waltenberger, C.-E.~Wulz\cmsAuthorMark{1}
% ...
\vskip\cmsinstskip
\dag:~Deceased\\
1:~~Also at Vienna University of Technology, Vienna, Austria\\
2:~~Also at CERN, European Organization for Nuclear Research, Geneva, Switzerland\\
% ...

\end{sloppypar}
\end{document}